# Magnetic Field Induced Phase Transition in Spinel GeNi$_2$O$_4$

T. Basu[1], T. Zou[1], Z. Dun[2], C. Q. Xu[1], C.R. Dela Cruz[3], Tao Hong[3], H.B. Cao[3], K.M. Taddei[3], H.D. Zhou[2], and X. Ke[1]

[1]*Department of Physics and Astronomy, Michigan State University, East Lansing, MI 48824, USA*

[2]*Department of Physics and Astronomy, University of Tennessee, Knoxville, TN 37996, USA*

[3]*Neutron Scattering Division, Oak Ridge National Laboratory, Oak Ridge, TN 37831, USA*

Cubic spinel GeNi$_2$O$_4$ exhibits intriguing magnetic properties with two successive antiferromagnetic phase transitions ($T_{N1} \approx 12.1$ and $T_{N2} \approx 11.4$ K) with the absence of any structural transition. We have performed detailed heat capacity and magnetic measurements in different crystallographic orientations. A new magnetic phase in presence of magnetic field ($H \geq 4$ T) along the [111] direction is revealed, which is not observed when the magnetic field is applied along the [100] and [110] directions. High field neutron powder diffraction measurements confirm such a change in magnetic phase, which could be ascribed to a spin reorientation in the presence of magnetic field. A strong magnetic anisotropy and competing magnetic interactions play a crucial role on the complex magnetic behavior in this cubic system.

I. **Introduction**

Geometrically frustrated magnetic systems with competing exchange interactions between magnetic ions placed on regular lattice tend to exhibit various exotic magnetic states, such as spin-glass, spin liquid, spin ice [1-6]. A small perturbation (e.g. external magnetic field, pressure) may release the magnetic frustration and often lead to a variety of unconventional long-range ordered state. Therefore, it is of high interest to understand the role of different key parameters in a material to tune such properties.

In the past two decades, there have been intense studies of spinel compounds, $AB_2O_4$. In particular, spinel compounds with magnetic ions occupying on B sites forming a pyrochlore lattice have attracted considerable attention due to the resultant magnetic frustrations [7]. When view along the <111> axis, the pyrochlore lattice consists of a stack of alternating Kagome and triangular planes. The germanium-based pyrochlore spinels, $Ge(TM)_2O_4$ (TM = transition metal ion, e.g. Co, Ni), exhibit quite distinct physical properties with contrasting magnetic behavior for different TM-ions [8-13]. The compound $GeCo_2O_4$ undergoes an antiferromagnetic (AFM) phase transition at $T_N \approx 21$ K [10, 13, 14], accompanied by a cubic-tetragonal structural transition relieving the magnetic frustration, similar to other spinels with half-integer spins such as (Zn, Cd, Hg)$Cr_2O_4$ ($S = 3/2$) [15-17]. In sharp contrast, $GeNi_2O_4$ shows two successive AFM orderings at $T_{N1} = 12.1$ K and $T_{N2} = 11.4$ K without any structural distortion, a feature that is rare and argued to be a consequence of the integer spin of $Ni^{2+}$ ($S = 1$) [8, 11, 13, 14, 18-20]. In addition, $GeCo_2O_4$ possesses a positive value ($\approx + 80$ K) of Curie-Weiss temperature ($\theta_W$), in contrast to the negative $\theta_W$ ($\approx - 15$ K) for $GeNi_2O_4$ [13].

By simply comparing the ratio of $T_N/\theta_W$, both $GeTM_2O_4$ compounds should not be regarded as highly frustrated magnets, where the magnetic frustration arises due to the competition of magnetic interactions between neighboring spins and further neighbors [13]. Neutron diffraction studies have shown that the ground state spin structure of these two compounds are similar, with a single ordering k-vector of (1/2 1/2 1/2), both having spins within the Kagome (triangular) planes ferromagnetically coupled whereas spins between the adjacent Kagome (triangular) planes being antiferromagnetically coupled [8, 13, 20, 21]. On the other hand, in contrast to $GeCo_2O_4$ in which spins in Kagome and triangular lattices order simultaneously at $T_N$ and have the same moment size, for $GeNi_2O_4$ spins in the Kagome planes partially order at $T_{N2} < T < T_{N1}$ but spins in the triangular planes remain completely disordered, which is followed by the ordering of spins in the latter planes at $T < T_{N2}$ [8, 13, 18, 20]. This feature implies that the magnetic interaction between adjacent Kagome and triangular planes is more substantial in $GeCo_2O_4$ than that in $GeNi_2O_4$. However, recent magnetic susceptibility and neutron diffraction measurements on $GeCo_2O_4$ have found two magnetic phase transitions occurring upon applying a magnetic field of ~ 4.6 T and 9.8 T [10], which are associated with the polarized spin alignment of triangular planes and Kagome planes [21], respectively. Strikingly surprising is that no such field-induced anomaly was reported on polycrystalline $GeNi_2O_4$ samples up to 14 T [11] and that two weak kinks in the $M$ ($H$) curves, which were ascribed to the spin reorientation, were seen at very high fields ($H$ = 30 T and 37 T) [13]. Obviously, this contrasting field dependent behavior in $GeCo_2O_4$ and $GeNi_2O_4$ could not reconcile with the fact that $GeNi_2O_4$ possesses lower values of $\theta_W$ and $T_N$ as well as two phase transition temperatures associated with the individual magnetic ordering of the Kagome and triangular planes compared to the former compound, which warrants further investigations.

In this paper, we report comprehensive heat capacity, magnetic susceptibility, powder neutron diffraction, and single crystal neutron diffraction measurements on $GeNi_2O_4$. We have observed a field-induced anomaly above a critical field (~ 4 T) applied along the [111] axis, which is absent for the field (up to 9 T) applied along the [100] and [110] axes as well as for polycrystalline samples. This new phase is presumably associated with the field-induced spin-reorientation.

## II. Experimental methods

Polycrystalline $GeNi_2O_4$ was synthesized using solid state chemistry method and the single crystal samples were grown using the floating zone technique under 5 atm $O_2$ atmosphere and with extra GeO2 added to compensate for the evaporation problem [22]. Laue x-ray diffraction was used to determine the crystallographic directions of single crystals and the samples were cut along the [100], [110] or [111] axes. Heat capacity and magnetic susceptibility measurements were performed using the adiabatic thermal relaxation technique and ACMS option on the Quantum Design PPMS cryostat, respectively. Single crystal neutron diffraction was performed at HB-3A four-circle diffractometer with a fixed wavelength of 1.003 Å in the High Flux Isotope Reactor (HFIR) at Oak Ridge National Laboratory (ORNL), and powder neutron diffraction measurements were carried out using both HB-2A powder diffractometer [23] and CG4C cold neutron triple axis spectrometer with the neutron wavelength 2.41 Å and 4.522 Å, respectively. The single crystal sample was cooled down using a closed-cycle cryostat on HB-3A, and the polycrystalline sample was cooled down using liquid helium cryostat with magnet on HB-2A and CG4C. The magnetic structural analysis is carried out using FullProf Suite [24] and SARAh [25] program.

## III. Results

The main panel of Fig. 1(a) shows the heat capacity data measured on the [111]-oriented $GeNi_2O_4$ single crystal at zero and 9 T magnetic field. At zero field, two sharp anomalies occurring at $T_{N1}$ = 12.1 K and $T_{N2}$ = 11.4 K were observed, which are presumably associated with two successive antiferromagnetic transitions in $GeNi_2O_4$ as reported in previous studies [8, 11, 13, 26].

In addition, there is a small but broad shoulder around 11 K, which has also been observed previously and speculated to be associated with some short-range correlation [11, 26]. Both transitions shift to lower temperature with increasing magnetic field, consistent with an antiferromagnetic order. Interestingly, upon applying a high magnetic field of 9 T along the [111] axis, there emerges a new anomaly at $T_{N3}^* = 10.7$ K, in addition to the existing two anomalies appearing at $T_{N1}^* = 11.8$ K and $T_{N2}^* = 9.6$ K which correspond to the original two magnetic phase transitions shifting to lower temperatures (to be discussed later). This field-induced new anomaly is absent for the polycrystalline sample, as shown in the inset of Fig. 1(a), consistent with the previous report [11].

Figure 1(b) presents the magnetic susceptibility data with 0.1 T and 9 T magnetic field applied along the [111] axis. In contrast to the earlier study by Hara et al where the transition was only observed at $T_{N2}$ with the field applied the [111] direction [26], here one can clearly see two kinks in the magnetic susceptibility data occurring at both $T_{N1}$ and $T_{N2}$ under 0.1 T field. In the presence of 9 T field, the sample exhibits three anomalies at $T_{N1}^*$, $T_{N2}^*$ and $T_{N3}^*$ which correspond nicely with the anomalies observed in the heat capacity measurements presented in Fig. 1(a), affirming the magnetic nature of all these phase transitions. The third anomaly at $T_{N3}^*$ is not observed when the field is applied along [100] or [110] directions or in the magnetic susceptibility measurement on polycrystalline sample (see Fig. S1, S2 and S3 in Supplemental Material [27]), which is consistent with the heat capacity results presented below.

To investigate the evolution of the field-induced new phase, heat capacity measurements were performed with various magnetic fields applied along the [111] axis, as shown in Fig. 1(c). Upon increasing the magnetic field, the temperature region between $T_{N1}^*$ and $T_{N2}^*$ enlarges with $T_{N2}^*$ decreasing while $T_{N1}^*$ remaining unchanged. When the field reaches about 4.0 T, another anomaly shows up, indicating the emergence of a new magnetic phase, while $T_{N1}^*$ gradually decreases. Again, such a field-induced new phase was not observed previously when measuring polycrystalline samples up to 14 T [11]. To further examine whether such a feature is universal or not, we carried out heat capacity measurement on GeNi$_2$O$_4$ single crystals with a 9 T magnetic field applied along the [100] and [110] directions as well. As shown in Fig. 1(d), no extra anomaly is observed in addition to a shift of $T_{N1}$ and $T_{N2}$ to lower temperatures with the field applied along these two directions. This indicates a strong magnetic anisotropy of GeNi$_2$O$_4$, which also accounts for the absence of new phases in the polycrystalline sample measured as shown in the inset of Fig. 1(a). Such an anisotropic behavior may originate from the higher order spin-orbit coupling that leads to a slight splitting between the spin-singlet excited state and the spin doublet ground state [8], in agreement with the small spin gap observed in the magnetic excitation measurements [11].

It is obvious that magnetism is quite complex with different competing magnetic interactions existing in this system, including the nearest and next-nearest exchange interactions between spins on triangular sites, the nearest and next-nearest exchange interactions between spins on Kagome sites, and the exchange interaction between spins on triangular and Kagome sites, etc., which can be largely influenced by external perturbations like magnetic field. To get a more detailed insight, we have performed isothermal dc magnetization up to 9 T at different temperatures around three magnetic transitions for $H$ // [111], as shown in Fig.2. The isothermal $M(H)$ is linear for $T > T_{N1}^*$ (say, $T = 12$ K), consistent with paramagnetic behavior. At $T = 11$ K ($T_{N3}^* < T < T_{N1}^*$), the $M(H)$ linearly increases with increasing $H$ up to 3.6 T, like AFM, then a weak step-like jump is observed around 3.9 T, and further linear variation is observed up to 9T. Such a sudden increase of magnetization as a function of field is attributable to $H$-induced magnetic transition, such as, spin-reorientation due to magnetic anisotropy, or, any meta-magnetic type transition. However, the *M*(*H*) does not tend to saturate and eventually the value of magnetization at 9 T reaches ~ 0.43 $\mu_B$, which is far less than its saturation value. Therefore, no transitions within our measurements can be attributed to polarizations of Ni moments. The *M*(*H*) at *T* = 10 K ($T_{N2}^* < T < T_{N3}^*$) exhibits a similar *H*-induced transition at further high magnetic field (*H* ~ 7.3 T). The increase of the transition field at 10 K compared to that measured at 11 K indicates an enhanced magnetic anisotropy at lower temperature. No clear *H*-induced magnetic transition up to 9 T is observed at further low temperature ($T \leq T_{N3}^*$), as shown in Fig.2 (i.e, *T* = 2 K and 9 K), which is attributed to development of strong magnetic anisotropy in this system below $T_{N3}^*$. Such an *H*-induced transition is absent in polycrystalline sample and also for H // [100] or [110]-directions (see inset of Fig. S1, S2 and S3 in Supplemental Material [27]).

To understand the magnetic structure associated with the field-induced new phase, neutron diffraction experiments have been carried out. First, we present zero-field diffraction data. Figure 3(a) shows the neutron powder diffraction data measured on HB-2A at *T* = 2 K and 25 K along with Rietveld fitting at 2 K. Consistent with the previous studies [8, 14, 18, 20], no structural transition at the onsets of magnetic phase transitions is observed and the material remains a cubic structure (space group: F d -3 m) down to the lowest temperature measured, unlike $GeCo_2O_4$ The magnetic Bragg peaks, such as (1/2 1/2 1/2) and (3/2 1/2 1/2), shown up at 2 K can be well indexed (see Fig.3a), in agreement with previous report [20]. An expanded view of magnetic Bragg peaks at low Q-value at 2 K compared to that of 25 K is depicted in Fig. S4 in the Supplementary Material [27]. Figure 3(b) shows the temperature dependence of magnetic Bragg peak intensity of these two magnetic Bragg peaks acquired on CG4C. One can clearly see two sharp increases in the intensity of both peaks upon cooling, which are associated with the magnetic phase transitions at $T_{N1}$ and $T_{N2}$.

We also performed single crystal neutron diffraction measurements using HB3A four-circle single crystal neutron diffractometer to collect various magnetic reflections. Figure 3(c,d) show plots of the comparison of observed and calculated intensities of various magnetic Bragg peaks collected at $T = 4.7$ K and $T = 11.6$ K in the absence of magnetic field. For the low temperature magnetic ground state, we obtain various possible magnetic structures with the same refinement quality. Spins in the adjacent Kagome and triangular planes could have two energetically equivalent arrangements. i) Spins in each individual triangular and Kagome planes are collinear in the (111) plane, but spins in the triangular plane could be either parallel (Fig. 4a) or perpendicular (Fig. 4b) to spins in the Kagome plane. Therefore, while spins in the Kagome plane are oriented along a certain direction (i.e., [1-10]), energetically spins in the triangular plane are oriented collinearly in any direction in the (111) plane. ii) Spins in the Kagome plane order collinearly in the (111) plane, whereas, spins in triangular plane order collinearly but with spins aligned perpendicularly to the (111) plane, as illustrated in Fig. 4(c). Note that spins in the neighboring Kagome planes and neighboring triangular planes are antiferromagnetically collinear to each other, as shown in Fig. 4. The spins in the Kagome plane have a moment of ~ 2.2 $\mu_B$/Ni while spins in the triangular plane has a reduced moment of 1.1 $\mu_B$/Ni, consistent with earlier reports [10, 17]. This, together with the energetically equivalent magnetic structures discussed above, indicates strong magnetic frustration of spins in the triangular plane, which arises from the competing exchange interactions between spins in a triangular plane and spins in its two neighboring Kagome planes. The data refinement at high temperature ($T_{N2} < T < T_{N1}$) gives similar spin structure in the Kagome plane with lower magnetic moment (Fig. 4d). However, the magnetic moment in the triangular plane is negligible, which suggests that spins in the triangular plane are disordered due to dominant magnetic frustration and thermal fluctuation [20]. The magnetic Bragg peak (1/2 1/2 1/2) does not vanishes completely at 25 K; instead it shows a weak but clear broadening peak compared to that measured at low temperature (see Fig. S4 in the Supplementary Material [27]), suggesting short-range magnetic correlation arising from magnetic frustration above long-range magnetic ordering.

Next, we present the neutron powder diffraction data measured in the presence of magnetic field. Figure 5(a) plots the neutron powder diffraction pattern measured at $T = 2$ K with $H = 0$ T and 6 T on HB-2A for comparison. An appreciable feature can be clearly observed: the intensity of the (1/2 1/2 1/2) Bragg peak is significantly enhanced, while the intensity of the (3/2 1/2 1/2) Bragg peak is drastically suppressed. A similar feature was observed previously in polycrystalline $GeCo_2O_4$ sample, which was ascribed to a combination of magnetic domain reorientation and a change of magnetic structure with either spins in neighboring triangular planes or spins in both neighboring triangular and Kagome planes turned into ferromagnetic alignment [21]. Nevertheless, distinct from $GeCo_2O_4$ where the nuclear Bragg peak intensity is also enhanced due to the field-induced canted ferromagnetic moment, no increase in nuclear Bragg peak intensity is observed in $GeNi_2O_4$ at 6 T, suggesting negligible ferromagnetic spin component at high field in either Kagome or triangular planes.

The temperature dependence of both (1/2 1/2 1/2) and (3/2 1/2 1/2) magnetic Bragg peak intensities measured at $H = 7$ T on CG4C is shown in Fig. 5(b). In contrast to that measured at zero field (shown in Fig. 3b), between at $T_{N1}^*$ and $T_{N3}^*$, the scattering intensity of the (3/2 1/2 1/2) Bragg peak increases below $T_{N1}^*$ and reaches a maximum at $T_{N3}^*$ while that of the (1/2 1/2 1/2) Bragg peak remains indistinguishable from the background signal. Interestingly, the scattering intensity of the (3/2 1/2 1/2) Bragg peak starts to decrease at $T_{N3}^*$ with lowering temperature and becomes nearly constant below $T_{N2}^*$ as shown in Fig. 5(b), in sharp contrast to the monotonic increases in the Bragg peak intensity observed at zero field. In the meantime, the scattering intensity of the (1/2 1/2 1/2) Bragg peak enhances below $T_{N3}^*$, exhibits a weak anomaly (change in slope) around $T_{N2}^*$ and continues to enhance with further lowering temperature. These features suggest that there are three magnetic phase transitions occurring upon applying a magnetic field, compared to two phase transitions observed at zero field, which is in agreement with the both heat capacity and magnetic susceptibility measurements presented in Fig. 1(a) and 1(b) respectively.

To better understand the field effect on the magnetic phase transitions, we have intended to refine neutron powder diffraction data measured in the presence of high magnetic field of 6T. The refinement results at 2 K for 6T are shown in Fig. S5 in the Supplemental Material [27]. All the magnetic peaks are well captured by same wave vector k = (1/2 1/2 1/2), although a slight mismatch in magnetic peak intensity at low angles is observed, which implies a similar magnetic structure to that measured at $H = 0$ T. The small mismatch in magnetic peak intensity at low angles between the observed and the calculated values is probably due to magnetic domain orientation of polycrystalline sample in the presence of high magnetic field. Figure 5(c) and 5(d) shows the magnetic field dependence of the scattering intensity of (1/2 1/2 1/2) and (3/2 1/2 1/2) magnetic Bragg peaks respectively, measured at $T = 1.5$ K upon increasing and decreasing the magnetic field. The observed hysteresis supports the scenario of magnetic domain orientations under high magnetic field, a feature similar to $GeCo_2O4$ reported in Ref. [18]. Such magnetic domain orientation after application of high magnetic field prevents from determining accurately the spin structure by refining the neutron diffraction data measured at 6 T. However, it is worth noting that sharp change in the scattering intensity of (1/2 1/2 1/2) and (3/2 1/2 1/2) magnetic Bragg peaks between $T_{N3}^*$ and $T_{N1}^*$ as a function of temperature shown in Fig. 5b cannot solely arise from domain orientations. Instead, there will be little magnetic domain orientation with increasing temperature once the magnetic domains are orientated at 2 K in the presence of 7 T magnetic field. As a result, the change in the magnetic Bragg peak intensity shown in Fig. 5b as a function of temperature is intrinsic and associated with the change in magnetic structure in the presence of magnetic field.

### IV. Discussions and conclusion

An unusual magnetic behavior is observed for GeNi$_2$O$_4$. The magnetic phase as a function of $T$ and $H$ is quite intriguing. In zero-magnetic field, the spins at Kagome sites are collinearly aligned along (111)-plane, while spins at triangular sites are quite energetically flexible in their arrangement due to magnetic frustration of exchange interactions between spins of each triangular plane and spins in the two adjacent Kagome planes. The spins in the triangular plane are aligned collinearly among themselves due to the internal field, but could be parallel or perpendicular to spins in the Kagome plane. An application of high magnetic field leads to distinct spin configurations. With a large magnetic field applied along the [111] direction, due to the addition of Zeeman interaction spins in the Kagome planes could not stabilize in (111) plane right below $T_{N1}^*$ unlike that of $H = 0$ T case. Instead, spins are aligned nearly along the external magnetic field, i.e., along the [111] direction. This is evidenced by the temperature dependence of ordering parameter measurements (Fig.5b), where the (1/2 1/2 1/2) magnetic peak intensity is negligible at $T_{N3}^* < T < T_{N1}^*$ but the (3/2 1/2 1/2) peak intensity increases with decreasing temperature. Note that neutrons couple to the magnetic moment perpendicular to the momentum transfer **q**, the observation of negligible intensity at (1/2 1/2 1/2) indicates that spins are aligned along the [111] direction; whereas, the non-zero neutron scattering intensity at (3/2 1/2 1/2) arises from the momentum transfer component perpendicular to the spin direction. Interestingly, below ~ $T_{N3}^*$ the (1/2 1/2 1/2) peak intensity start to increase sharply and consequentially (3/2 1/2 1/2) peak intensity decreases (Fig.5b). This implies that there is spin-reorientation of Kagome spins from nearly the [111] axis to the (111)-plane below $T_{N3}^*$. Based on the zero-field spin structure shown in Fig. 4(a,b,c), it suggests that the anisotropy of spins in the triangular plane is weak as spins could be aligned either in the (111) plane or perpendicular to the (111) plane, while Kagome plane spins exhibit strong magnetic anisotropy with spins fixed in the (111) plane. It is likely that the magnetic anisotropy gets enhanced at low temperature which facilitates to align spins in the easy-plane despite of strong magnetic field. The weak transition around 3.9 T in *M*(*H*) at 11 K (where only Kagome spins order) and the negligible ferromagnetic spin component in neutron diffraction data measured at 11.3 K under 6 T are consistent with spin-reorientation phenomenon of Kagome spins in this system. Both heat capacity and magnetic susceptibility (Fig. 1(a,b)) exhibit another anomaly at $T_{N3}^*$ for high magnetic field (*H* > 3T) where the spin-orientation occurs. Below $T_{N2}^*$ both the (1/2 1/2 1/2) and (3/2 1/2 1/2) peak intensity starts to increase slightly with further lowering temperature (Fig. 5b). Correspondingly, an anomaly is observed in both heat capacity and magnetic susceptibility measurements. We speculate that below $T_{N2}^*$ spins in the triangular plane are ordered. Because of powder averaging effect and magnetic domain orientation in neutron powder diffraction measurements in the presence of high magnetic field, single crystal neutron diffraction is warranted in order to unequivocally unravel the field-induced spin structure.

To briefly summarize the *T* and *H* dependence of magnetic phases, a phase diagram is illustrated in Fig. 6 where the data points are taken from heat capacity measurements with *H* applied along the [111] crystalline orientation. A long-range collinear magnetic ordering is observed below $T_{N2}^*$ (stripe region below red curve in Fig.6) where spins on both Kagome and triangular plane orders. The magnetic ordering temperature decreases sequentially with an application of magnetic field, consistent with the AFM nature. Above $T_{N2}^*$ (above red curve in Fig.6), spins in the triangular plane are disordered and only spin ordering in the Kagome plane is observed below $T_{N1}^*$. The spins on Kagome plane order collinearly in the (111)-plane below 12.1 K ($T_{N1}^*$) up to 3 T, and a strong spin-reorientation from the (111)-plane to its perpendicular direction is observed above ~ 3 T. The yellow shaded region in Fig.6 (i.e. $T_{N3}^* < T < T_{N1}^*$) depicts the arrangement of Kagome spins which order perpendicular to the (111)-plane. However, the (111)-collinear arrangement persists at lower temperature ($T_{N2}^* < T < T_{N3}^*$) even above 3 T, which is illustrated in the light blue shaded region.

The field-induced phase transition is not observed with the magnetic field applied along (100) and (110) crystalline orientations as well as in polycrystalline samples, as evidenced from the magnetic susceptibility, isothermal magnetization, and heat capacity measurements (see Fig. 1(a,d), Fig. 2, and Fig. S1, S2 and S3 in Supplemental Material [27]). When magnetic field is applied along the [100] or [110] direction (see Fig.1d), no field-induced spin-reorientation is observed, because the magnetic field favors such spin-arrangement (nearly perpendicular to hard-axis). Therefore, the Kagome spins remain colinear in the (111)-plane throughout the magnetic field and temperature range below $T_{N1}^*$. It is worth noting that all single crystals measured, including (100), (110), and (111)-oriented, were cut from the same big piece of single crystal, therefore, it is reasonable to argue that these crystal have the same crystal quality. Further, we have grounded the measured (111)-oriented single crystal and remeasured the magnetic susceptibility of thus-formed polycrystalline sample (see Fig.S3 Supplemental Material [27]) which does not yield any magnetic anomaly related to $T_{N3}^*$). Therefore, such an unusual anisotropic field-induced magnetic behavior in GeNi$_2$O$_4$ compound is attributable to intrinsic magnetic anisotropy which may arise from higher-order spin-orbit coupling. This cubic spinel consists of NiO$_6$ octahedra, as a result $Ni^{+2}$ -magnetic ground state splits in an octahedral crystal field. However, the $Ni^{2+}$ ions are located on the vertices of corner sharing tetrahedra. It was proposed in Ref. [8] that because of this trigonal crystal-field effect and second-order spin-orbit coupling, the ground state exhibits further splitting. The Lande-g factor of ~2.33 instead of spin-only value (~2) further supports the spin-orbit coupling in this system [8]. Such spin-orbit coupling may be responsible for the observed spin excitation gap [11], consistent with the magnetic anisotropy character that we present in this work.

In summary, we have disclosed a new phase in spinel $GeNi_2O_4$ driven by magnetic field applied along [111] crystallographic direction. In zero-magnetic field spins at Kagome site are collinearly align along (111)-plane and spins at triangular site are quite energetically flexible in their arrangement due to magnetic frustration. Application of magnetic field along the [111] direction aligns spins in the Kagome plane along the magnetic field at high temperature while spins in the triangular plane remains disordered. Upon lowering the temperature, another phase transition occurs with spins in the Kagome plane reoriented to the (111)-plane due to the enhanced magnetic anisotropy. Further decreasing temperature eventually leads to the ordering of spins in the triangular plane. Such a *T-H* dependence of spin configuration in $GeNi_2O_4$ results from the synergetic effect of magnetic anisotropy, Zeeman interaction, and competing exchange interactions.

**Acknowledgements**

T.B. and X.K. thank the financial support by the U.S. Department of Energy, Office of Science, Office of Basic Energy Sciences, Materials Sciences and Engineering Division under Award # DE-SC0019259. T.Z. was supported by the Start-up funds at Michigan State University.

Z. D. and H. Z. thank the financial support from U.S. Department of Energy, Office of Science, Office of Basic Energy Sciences, Materials Sciences and Engineering Division under Award # DE-SC0020254. A portion of this research used resources at the High Flux Isotope Reactor, a DOE Office of Science User Facility operated by the Oak Ridge National Laboratory.

Figure captions:

Figure 1: (a) The main panel displays the temperature variation of heat capacity of single crystal GeNi$_2$O$_4$ measured under zero magnetic field and with 9 T magnetic field applied along the [111] crystallographic axis. The inset shows heat capacity measurement acquired on polycrystalline GeNi$_2$O$_4$ samples. (b) Magnetic susceptibility of single crystal GeNi$_2$O$_4$ measured with 0.1 and 9 T magnetic field applied along the [111] axis, respectively. (c) Temperature dependence of heat capacity of single crystal GeNi$_2$O$_4$ with various fields applied along the [111] axis. The absolute value along y-axis is slightly shifted to get a better visualization of peak shifting for different field. (d) Temperature dependence of heat capacity of single crystal GeNi$_2$O$_4$ with 9 T magnetic field applied along the [100], [110] and [111] axes, respectively.

Figure 2: Isothermal magnetization at selective temperature of GeNi$_2$O$_4$ single crystal when magnetic field is applied along the [111] crystallographic axis.

Figure 3: Powder neutron diffraction pattern of GeNi$_2$O$_4$ collected at 2 K and 25 K under zero field. The black and green symbols represent the experimental data, while the red solid line shows the Rietveld fitting for 2K neutron data. The vertical bars display the Bragg peak positions. The upper vertical lines represents Bragg peaks of crystal structure GeNi$_2$O$_4$, the middle vertical lines represents Bragg peaks of Al sample can, the lower vertical line represents magnetic Bragg peaks of GeNi$_2$O$_4$. The magenta curve shows the difference between the experimental and calculated intensity for 2 K data. (b) Peak intensity of (1/2 1/2 1/2) and (3/2 1/2 1/2) magnetic Bragg peaks as function of temperature under zero field. (c) and (d) best fit of Rietveld refinement of single crystal data at 4.7 K and 11.6 K, respectively.

Figure 4: Possible spin structures (a), (b) and (c) at 4.7 K (below $T_{N2}$) and (d) 11.6 K (below $T_{N1}$) in the absence of applied magnetic field, as described in the text.

Figure 5: (a) Powder neutron diffraction of $GeNi_2O_4$ collected at 2K under $H = 0$ T and 6 T magnetic field. (b) Peak intensity of (1/2 1/2 1/2) and (3/2 1/2 1/2) magnetic Bragg peaks as a function of temperature under 6 T magnetic field. (c) and (d) Magnetic field dependence of neutron scattering intensity at 2 K of (1/2 1/2 1/2) and (3/2 1/2 1/2) Bragg peaks respectively, in warming (open symbol) and cooling (closed symbol) of field condition.

Figure 6. *T-H* phase diagram with the magnetic field applied along the [111] crystalline orientation to visualize the change in magnetic structure as a function of *T* and *H*.

Figure 1

T. Basu et al,

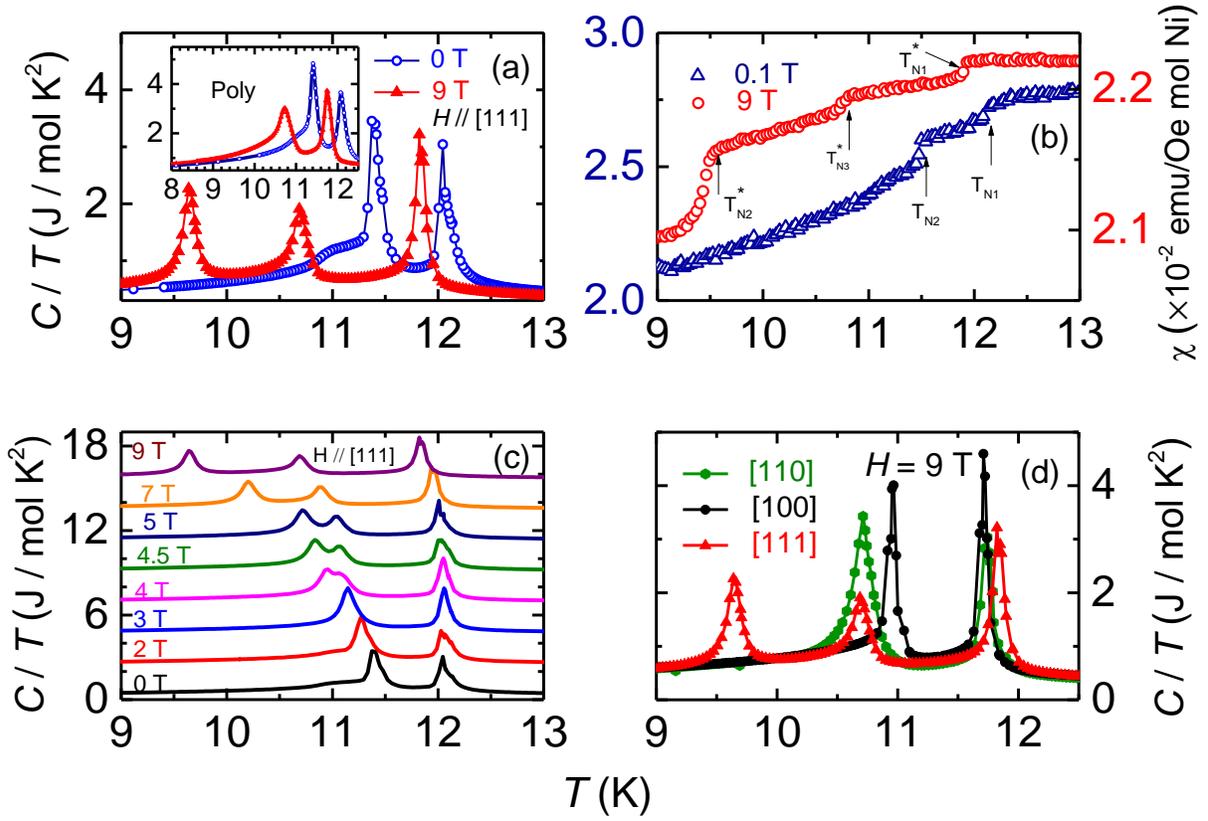

Figure 2

T. Basu et al,

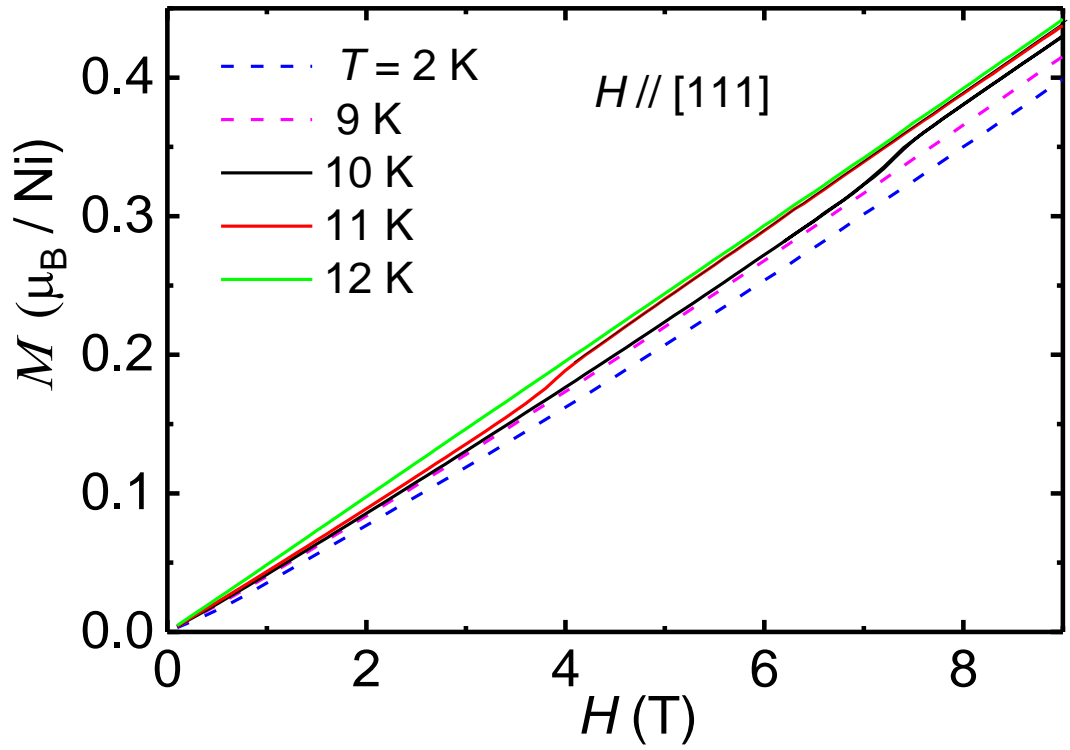

Figure 3

T. Basu et al,

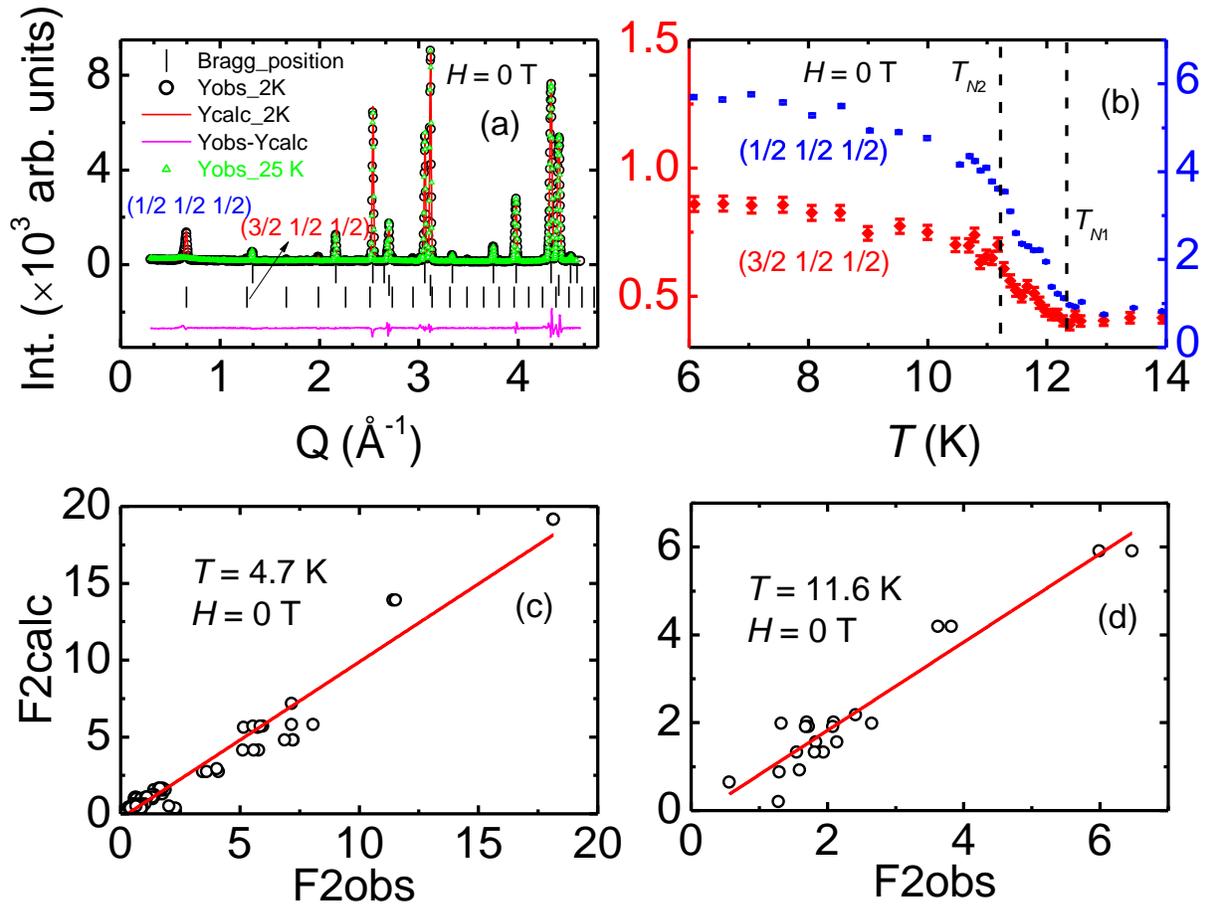

Figure 4

T. Basu et al,

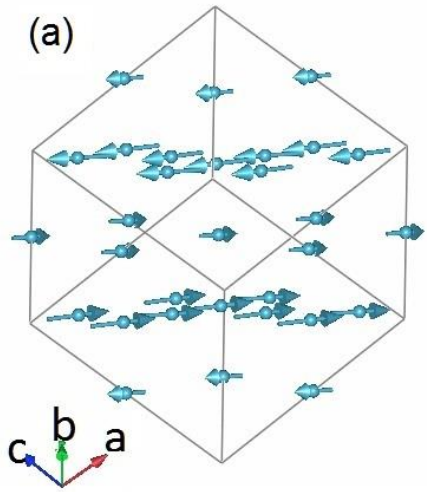
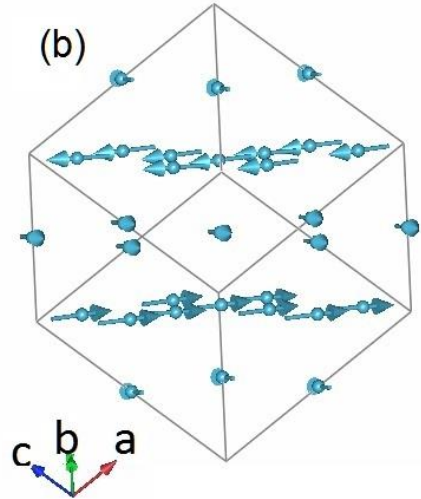
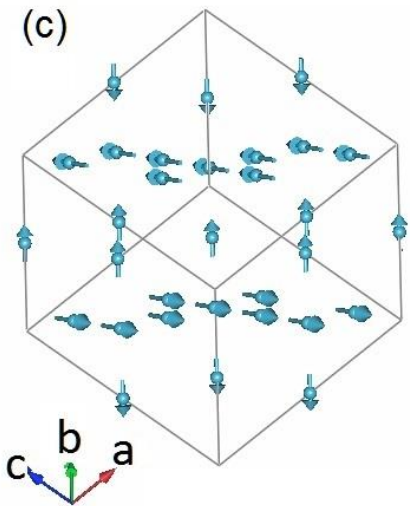
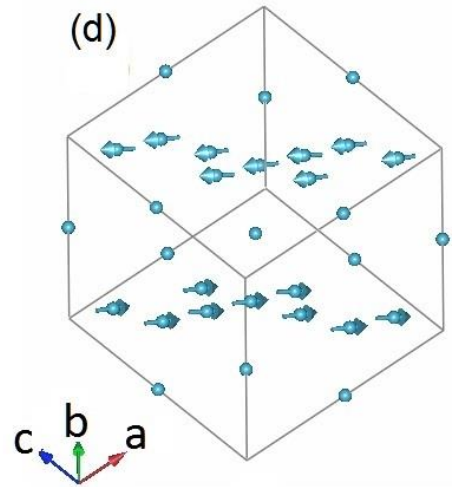

Figure 5

T. Basu et al,

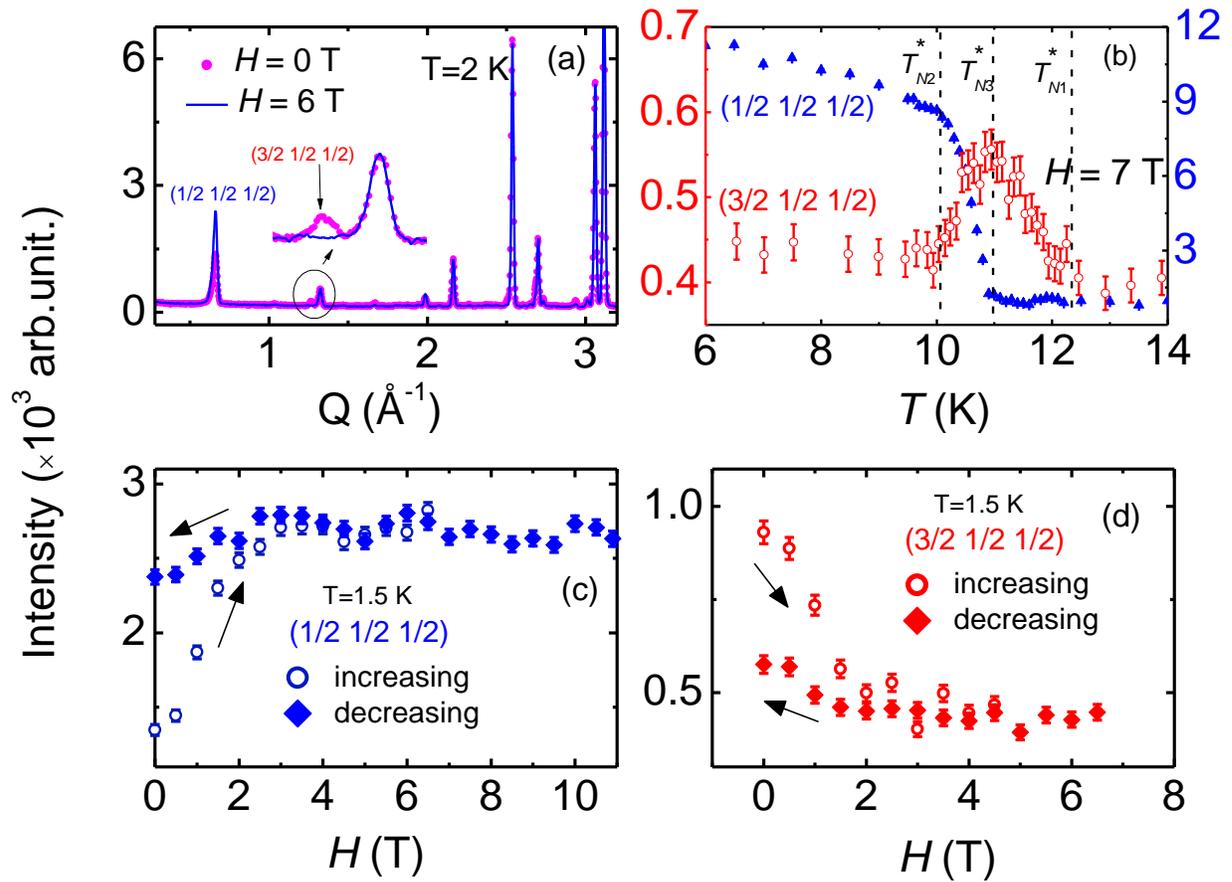

Figure. 6

T. Basu et al,

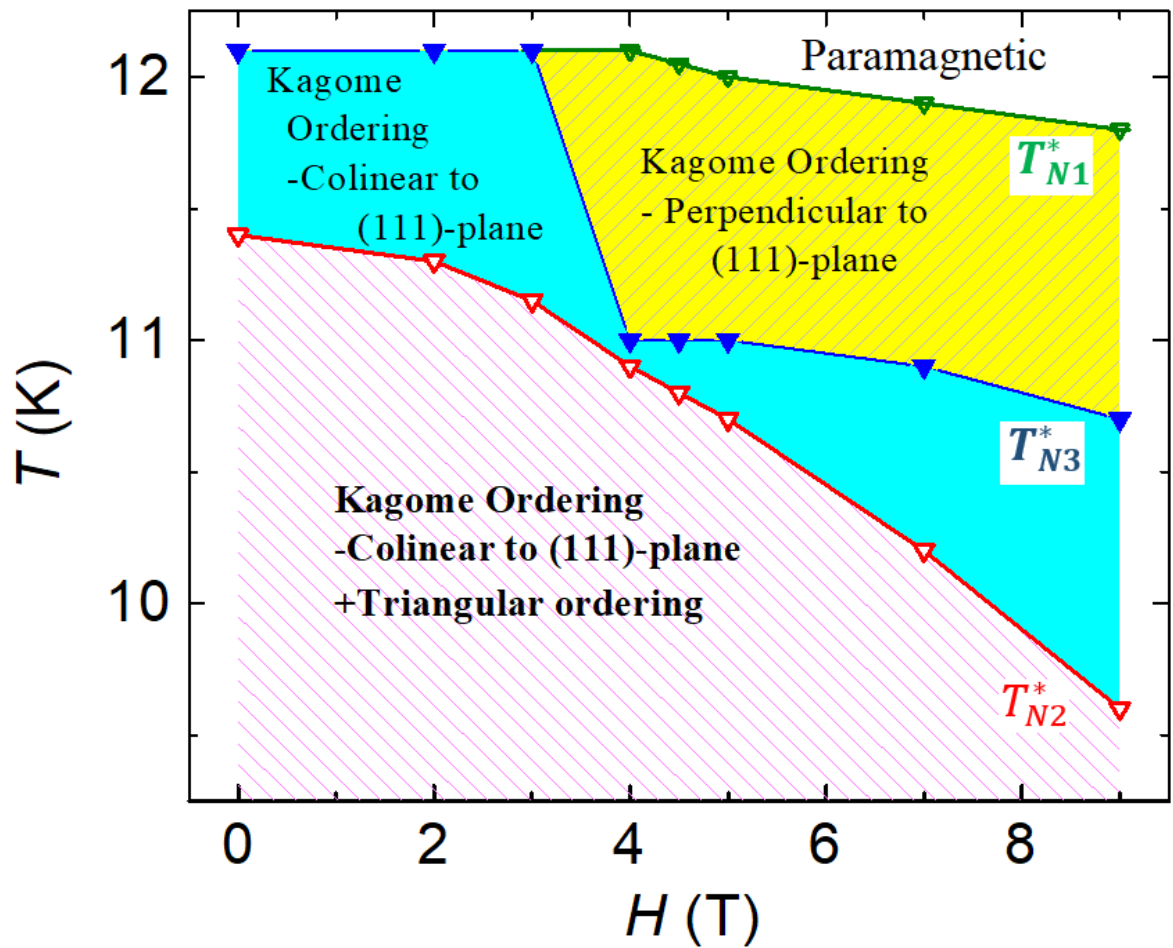

# Supplementary Material

## A. Magnetization for [100], [110] and polycrystalline sample of $GeNi_2O_4$:

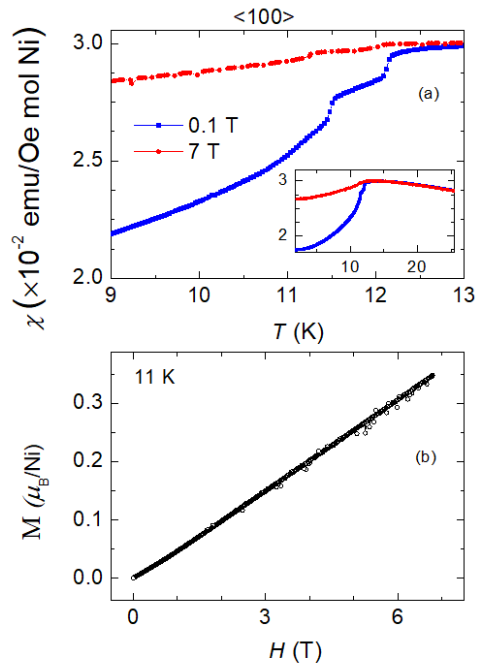

**Figure S1:** (a) Temperature dependent Magnetic susceptibility of single crystal $GeNi_2O_4$ measured with 0.1 and 7 T magnetic field applied along the [100] axis from 9-13 K; Inset: shows the broad *T*-range. (b) Isothermal magnetization at 11 K.

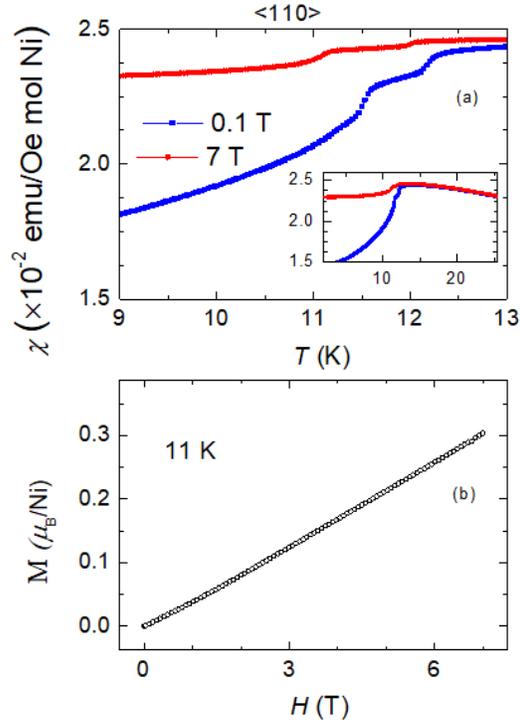

**Figure S2:** Temperature dependent Magnetic susceptibility of single crystal GeNi$_2$O$_4$ measured with 0.1 and 7T magnetic field applied along the [110] axis from 9-13 K; Inset: shows the broad *T*-range. (b) Isothermal magnetization at 11 K.

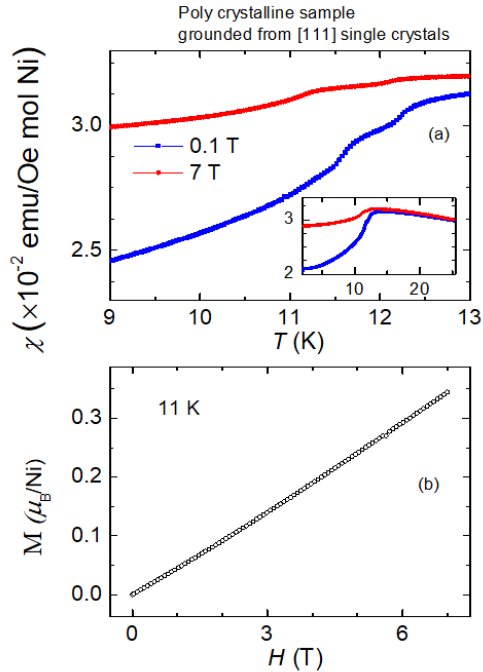

**Figure S3:** Temperature dependent Magnetic susceptibility of polycrystalline sample (which is grounded from [111] single crystals specimen) measured with 0.1 and 7 T magnetic field applied along the [100] axis from 9-13 K; Inset: shows the broad *T*-range. (b) Isothermal magnetization at 11 K.

## B. Neutron diffraction

The magnetic structure is resolved using both SARAh [1] and FullProf Suite [2] program. The compound GeNi$_2$O$_4$ crystalizes in space group Fd-3m (227). The magnetic atom Ni (0.5, 0.5, 0.5) with ordering wave vector k=0 0 0 gives rise to two inequivalent magnetic atomic sites, Ni1 (0.5, 0.5, 0.5) and Ni2 (0.25, 0.25, 0.5). There are two irreducible representations ($\Gamma_2$ and $\Gamma_6$) associated with Ni1-site and two irreducible representations ($\Gamma_3$ and $\Gamma_5$) associated with Ni2-site, as described in Table-1 and Table-2 respectively, obtained from SARAh program [1]. A mixing of these irreducible representations gives the best refinement results.

| IR | BV | Atom | BV components ||||||
|---|---|---|---|---|---|---|---|---|
| | | | $m_{\|a}$ | $m_{\|b}$ | $m_{\|c}$ | $im_{\|a}$ | $im_{\|b}$ | $im_{\|c}$ |
| $\Gamma_2$ | $\psi_1$ | 1 | 12 | 12 | 12 | 0 | 0 | 0 |
| $\Gamma_6$ | $\psi_2$ | 1 | 6 | 6 | -12 | 0 | 0 | 0 |
| | $\psi_3$ | 1 | -10.392 | 10.392 | 0 | 0 | 0 | 0 |

Table 1: Basis vectors for the space group F d -3 m:2 with **k** = (0.5, 0.5, 0.5). The decomposition of the magnetic representation for the Ni1 site (0.5, 0.5, 0.5) is $\Gamma_{\text{Mag}} = 0\Gamma_1^1 + 1\Gamma_2^1 + 0\Gamma_3^1 + 0\Gamma_4^1 + 0\Gamma_5^2 + 1\Gamma_6^2$. The atom of the primitive basis is defined according to 1: (0.5, 0.5, 0.5).

| IR | BV | Atom | BV components ||||||
|---|---|---|---|---|---|---|---|---|
| | | | $m_{\|a}$ | $m_{\|b}$ | $m_{\|c}$ | $im_{\|a}$ | $im_{\|b}$ | $im_{\|c}$ |
| $\Gamma_3$ | $\psi_1$ | 1 | 4 | 4 | 4 | 0 | 0 | 0 |
| | | 2 | 4 | 4 | 4 | 0 | 0 | 0 |
| | | 3 | 4 | 4 | 4 | 0 | 0 | 0 |
| | $\psi_2$ | 1 | 4 | 4 | -8 | 0 | 0 | 0 |
| | | 2 | 4 | -8 | 4 | 0 | 0 | 0 |
| | | 3 | -8 | 4 | 4 | 0 | 0 | 0 |
| $\Gamma_5$ | $\psi_3$ | 1 | 4 | -4 | 0 | 0 | 0 | 0 |
| | | 2 | 2 | 0 | -2 | 0 | 0 | 0 |
| | | 3 | 0 | -2 | 2 | 0 | 0 | 0 |
| | $\psi_4$ | 1 | 0 | 0 | 0 | 0 | 0 | 0 |
| | | 2 | 3.464 | 3.464 | 3.464 | 0 | 0 | 0 |
| | | 3 | -3.464 | -3.464 | -3.464 | 0 | 0 | 0 |
| | $\psi_5$ | 1 | 0 | 0 | 0 | 0 | 0 | 0 |
| | | 2 | 3.464 | -6.928 | 3.464 | 0 | 0 | 0 |
| | | 3 | 6.928 | -3.464 | -3.464 | 0 | 0 | 0 |
| | $\psi_6$ | 1 | 0 | 0 | 0 | 0 | 0 | 0 |
| | | 2 | 3.464 | 0 | -3.464 | 0 | 0 | 0 |
| | | 3 | 0 | 3.464 | -3.464 | 0 | 0 | 0 |
| | $\psi_7$ | 1 | 4 | 4 | 4 | 0 | 0 | 0 |
| | | 2 | -2 | -2 | -2 | 0 | 0 | 0 |
| | | 3 | -2 | -2 | -2 | 0 | 0 | 0 |
| | $\psi_8$ | 1 | 4 | 4 | -8 | 0 | 0 | 0 |
| | | 2 | -2 | 4 | -2 | 0 | 0 | 0 |
| | | 3 | 4 | -2 | -2 | 0 | 0 | 0 |

Table 2: Basis vectors for the space group F d -3 m:2 with **k** = (0.5, 0.5, 0.5). The decomposition of the magnetic representation for the Ni2 site (0.25,0 2.5, .5) is $\Gamma_{\text{Mag}} = 0\Gamma_1^1 + 0\Gamma_2^1 + 2\Gamma_3^1 + 0\Gamma_4^1 + 3\Gamma_5^2 + 0\Gamma_6^2$. The atom of the primitive basis is defined according to 1: (0.25, 0.25, 0.5), 2: (0.25, 0.5, 0.25), 3: (0.5, 0.25, 0.25).

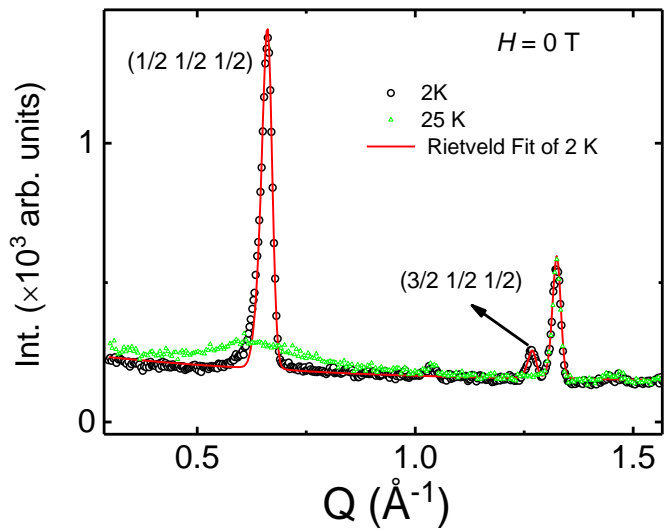

**Figure S4:** Powder neutron diffraction pattern of GeNi$_2$O$_4$ collected at 2 K and 25 K under zero field at low Q-value. The magnetic Bragg peaks (1/2 1/2 1/2) and (3/2 1/2 1/2) are highlighted.

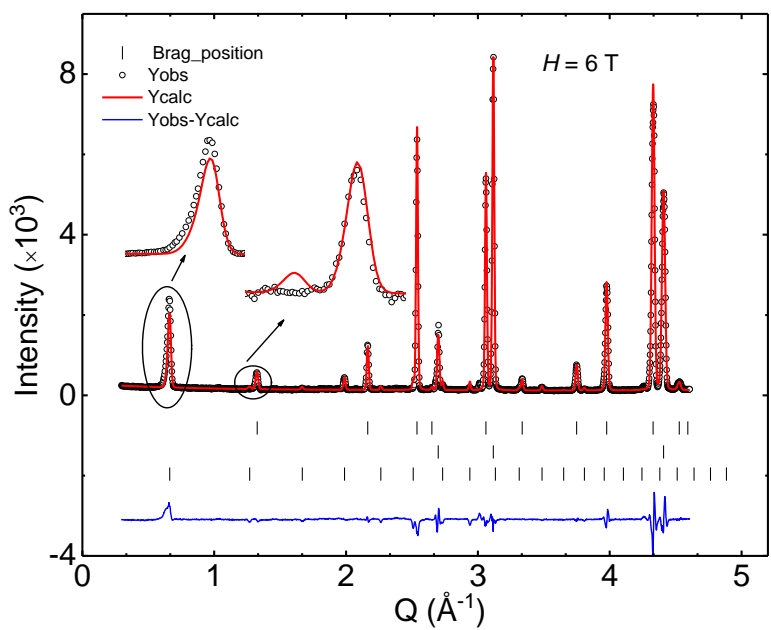

**Figure S5:** Powder neutron diffraction pattern of GeNi$_2$O$_4$ collected at 2 K under 6T magnetic field. The black symbols represent the experimental data, while the red solid line shows the Rietveld fitting for neutron data. The vertical bars display the Bragg peak positions. The upper vertical lines represents Bragg peaks of crystal structure GeNi$_2$O$_4$, the middle vertical lines represents Bragg peaks of Al sample can, and the lower vertical line represents magnetic Bragg peaks of GeNi$_2$O$_4$ for wave vector k = (1/2 1/2 1/2). The magenta curve shows the difference between the experimental and calculated intensity. A clear view of magnetic Bragg peak (1/2 1/2 1/2) and (3/2 1/2 1/2) are highlighted in magnified plot as inset.

Whereas the change in relative peak intensity between $H = 0$ T and 6 T may indicate a change in magnetic structure, we do not find any other better solution in 6 T compared to that of $H = 0$ T. The mismatch between calculated and experimental intensity is presumably due to the domain orientations in the presence of high magnetic field at 2 K, which is discussed in the main text. In addition to the collinear spin structure in the Kagome plane, we have also checked if any canted AFM spin structure is more plausible. However, the Rietveld refinement gives similar fitting results. We have also checked the possibility of ferromagnetic moment on the triangular sites, which does not fitted well the experimental data either. Therefore, we cannot conclude whether any canted magnetic structure is possible or not from the neutron powder data. Single crystal neutron diffraction measurements under 6 T are warranted to unequivocally determine the field-induced spin structure.